\documentclass[twocolumn]{aastex631}
\usepackage{amsmath}
\usepackage{color}
\usepackage{natbib}
\usepackage{amssymb}
\usepackage{amsfonts}
\usepackage{stackengine}
\usepackage{changepage}
\usepackage{subfigure}
\usepackage{graphicx}
\usepackage{rotating}
\usepackage{threeparttable}
\usepackage{xspace}
\usepackage{soul}
\newcommand{\beq}{\begin{equation}}
\newcommand{\eeq}{\end{equation}}
\newcommand{\g}{\gamma}
\def\ba{\begin{eqnarray}}
\def\ea{\end{eqnarray}}

\shorttitle{ }

\shortauthors{Peng et al.}
\begin{document}

\title{GRBs from Collapse of Thorne-Żytkow-like Objects as the Aftermath of WD-NS Coalescence}

\correspondingauthor{Zong-kai Peng}
\email{zkpeng@bnu.edu.cn}
\correspondingauthor{Bin-Bin Zhang}
\email{bbzhang@nju.edu.cn}
\correspondingauthor{He Gao}
\email{gaohe@bnu.edu.cn}

\author[0000-0003-1904-0574]{Zong-kai Peng}
\affiliation{Institute for Frontier in Astronomy and Astrophysics, Beijing Normal University, Beijing 102206, China}
\affiliation{Department of Astronomy, Beijing Normal University, Beijing 100875, China}
\affiliation{School of Astronomy and Space Science, Nanjing
University, Nanjing 210093, China}

\author[0000-0002-5550-4017]{Zi-ke Liu}
\affiliation{INPAC, Shanghai Key Laboratory for Particle Physics and Cosmology, School of
Physics and Astronomy, Shanghai Jiao-Tong University, Shanghai 200240, China}

\author[0000-0003-4111-5958]{Bin-Bin Zhang}
\affiliation{School of Astronomy and Space Science, Nanjing
University, Nanjing 210093, China}
\affiliation{Key Laboratory of Modern Astronomy and Astrophysics (Nanjing University), Ministry of Education, China}

\author[0000-0003-2516-6288]{He Gao}
\affiliation{Institute for Frontier in Astronomy and Astrophysics, Beijing Normal University, Beijing 102206, China}
\affiliation{Department of Astronomy, Beijing Normal University, Beijing 100875, China}

\begin{abstract}

The Type II gamma-ray burst (GRB) 200826A challenges collapsar models by questioning how they can generate a genuinely short-duration event. The other Type I GRB 211211A confused us with a kilonova signature observed in the afterglow of a \textit{long} burst. Here, we propose a comprehensive model in which both bursts are the results of the collapse of Thorne-Żytkow-like Objects (TZlOs). The TZlO consists of a central neutron star (NS), with a dense white dwarf (WD) material envelope, which is formed as the aftermath of a WD-NS coalescence.  We find that the characteristics of the resultant GRBs depend on whether the TZlO collapses immediately following the WD-NS merger or not. Additionally, the observational properties of the consequent GRBs manifest variations contingent upon whether the collapse of the TZlO results in a magnetar or a black hole. We also show that our model is consistent with the observations of GRB 211211A and GRB 200826A.  Specifically, the optical excess in GRB 211211A can be attributed to an engine-fed kilonova, while the \textit{supernova bump} in GRB 200826A is likely due to the collision between the ejecta and the disk wind shell. 

\end{abstract}

\section{Introduction}

The long-standing two-type origins \citep{Zha2011} of gamma-ray bursts (GRBs) have recently been challenged by some newly discovered observations, which added that a long GRB could originate from a merger-triggered core collapse \citep{Tho2011,Yang2022}, whereas a short GRB can originate from a magnetar giant flare \citep[e.g., GRB 200415A;][]{Yan2020} or an unusual collapse of a massive star \citep[e.g., GRB 200826A;][]{ Ahu2021, Zha2021}. In particular, \cite{Zha2021} suggested a possibility that the intrinsically short duration of the Type II GRB 200826A can be explained by a progenitor involved with a compact object, such as a white dwarf (WD), which supplies much denser materials to account for the short-accretion timescale to match the observed short duration. However, as pointed out by \cite{Zha2021}, isolated WDs are incapable of producing GRBs; one, therefore, has to invoke a merger process between a WD and one other compact star such as a neutron star (NS) or a black hole (BH). One such combination, which has gained increasing interest in the field, is the WD-NS merger. The WD-involved binaries have long been proposed to serve as GRB central engines \citep{Bel2002, Mid2004}; however, less attention has been paid to how they generate short-duration GRBs. Furthermore, \citet{Yang2022} investigated the long GRB 211211A, characterized by extended emission (EE) and the detection of a kilonova signal during its afterglow phase. Contrary to the conventional understanding that long bursts originate from the collapse of massive stars, \citet{Yang2022} propose that this particular burst can be attributed to the merger of compact objects. \citet{Yang2022} argue that the WD-NS merger provides a compelling explanation for the observed features of this burst. The second-brightest gamma-ray burst, GRB 230307A, shares some characteristics with GRB 211211A, such as both are long GRBs with notable excesses in their optical afterglows \citep{Sun2023, Yang2024}, suggesting that GRB 230307A might also be a product of the WD-NS merger \citep{Wang2024}.

A WD-NS system can follow two paths, depending on the mass ratio between the WD and NS, i.e., $q = M_{\rm WD}/M_{\rm NS}$. For $q < q_{\rm crit}$ ($q_{\rm crit}$, the critical mass ratio), the WD will fill the Roche lobe slowly and undergo stable mass transfer (SMT). In this case, the WD-NS system will form an ultracompact X-ray binary. For $q > q_{\rm crit}$, the WD will be disrupted by the NS and will suffer unstable mass transfer\footnote{The value of $q_{\rm crit}$ is related to the composition of the progenitor stars as well as the process of accretion disk formation. The results from \citet{Ver1988} indicate $q_{\rm crit} \approx 0.5$. \citet{Pas2009} provided a broader range, suggesting $q_{\rm crit}$ to be in the range of $\sim 0.2-0.5$. They proposed that for $q > q_{\rm crit,max}$, the process is characterized as UMT, for $q < q_{\rm crit,min}$ as SMT, and the process remains uncertain for $q_{\rm crit,min} <q < q_{\rm crit,max}$. The results of \citet{Bob2017} suggest that within WD-NS systems, only helium WD with a mass less than $0.2~M_{\odot}$ undergo the SMT process, while in other scenarios, the occurrence of mass transfer is typically characterized as UMT.} \citep[UMT;][]{Pas2011}. Within the framework of the SMT process, it is difficult to envisage a connection with intense astrophysical burst events like GRBs. However, the WD-NS system, through the UMT process, could potentially result in some violent bursting events \citep{Yang2022, Zhong2023}. During the disruption process, the WD is stretched by the tidal force, producing self-interaction shocks, and circularizing the material into a disk\citep{Pas2011, Zen2020, Bob2022}. In this phase, the temperature and density in the middle region of the disk are high and may cause nuclear burning. \citet{Mar2016} and \citet{Zen2019} did simulations of the nuclear burning process, and their results showed that $10^{-4}-10^{-3}~ M_{\odot}$ of $^{56}\rm Ni$ was formed. \citet{Met2012}, \citet{Fer2019}, \citet{Zen2020}, and \citet{Kal2023} provided higher values around $10^{-3} -10^{-2}~ M_{\odot}$. \citet{Bob2022} pointed out that more $^{56}\rm Ni$ up to $0.05 -0.1~ M_{\odot}$ was synthesized during the process of massive ONe WD-NS mergers. Contemporary numerical simulations have not yet reached a consensus on the quantity of material expelled in the aftermath of WD-NS mergers. By synthesizing the outcomes of numerical analyses conducted by \citet{Fer2019}, \citet{Zen2019}, and \citet{Kal2023}, it is deduced that approximately $10-80\%$ of the initial WD mass is ejected. This outflow, which includes small amounts of $^{56}$Ni, gives rise to supernova (SN)-like transients characterized by a rapid evolution and diminished luminosity. The nuclear reaction processes, characteristics of the disk wind, and the nature of transients emanating from these mergers have been extensively discussed by \citet{Bob2017}, \citet{Zen2019, Zen2020}, \citet{Bob2022}, and \citet{Kal2023}. These studies propose that certain atypical transient phenomena, such as Ca-rich transients, faint Type Iax supernovae (SNe), and some faint rapid red transients, may be attributable to WD-NS merger events.

In the case that the mass of WD approaches the Chandrasekhar limit, the orbital period of the WD-NS system becomes very short when the UMT phase starts. In such instances, the NS may not be able to completely disrupt the WD but may plunge into the WD \citep{Yang2022}. After the merger process, the remnant consists of a central NS, a dense WD material envelope, and a disk. This type of celestial body, where an NS occupies the core of a WD, bears a structural resemblance to Thorne-Żytkow Objects \citep[TZOs;][]{Tho1977}. We refer to these as TZlOs \citep{Pas2011} in our context. These objects exist in a super-Chandrasekhar limit state, and their eventual collapse may potentially trigger the production of GRBs \citep{Yang2022}. The GRBs originating from such progenitors exhibit distinct properties in their prompt emissions and subsequent afterglow emissions, differing from those typically associated with the collapse of massive stars or BH/NS-NS mergers.

In this paper, we propose that the collapse of a TZlO, as the aftermath of a WD-NS merger, can be one of the progenitors of GRBs. We present our physical picture in \S 2 and derive our model's observable components as a fittable model in \S 3. We then apply our model to the interpretation of two peculiar GRBs in \S 4. A brief summary and discussion follow in \S5.

\section{The Physical Picture}

\citet{Pas2011} conducted general relativistic hydrodynamics simulations of the WD-NS merger process, with their results suggesting that the post-merger product can be described as a TZlO surrounded by a massive disk system. Subsequent studies by \citet{Met2012}, \citet{Mar2016}, \citet{Bob2017}, \citet{Fer2019}, \citet{Zen2019, Zen2020}, \citet{Bob2022} and \citet{Kal2023} also conducted numerical simulations of the WD-NS merger process, yet their findings did not mention the formation of TZlO objects. Although \citet{Zen2020} posited that the leftover WD material would fall back and envelop the NS to form a spherical structure after the disk evolution concludes, it did not describe this product as a TZlO. Predominantly, these numerical simulation efforts on WD-NS mergers primarily focus on the nucleosynthesis process and the observational effects of outflows produced by disk winds, with the post-merger system's late-stage evolution still requiring further research. Additionally, current numerical simulations have not yet addressed mergers involving WDs and NSs of extreme masses, which represents a future avenue for simulation efforts. Drawing on the physical description of the post-merger product by \citet{Pas2011}, we speculate that the merger of a WD near the Chandrasekhar limit with an NS would culminate in a system comprising a TZlO and a disk. The collapse of the TZlO's envelope may potentially lead to GRBs.

The formation and collapse of a TZlO is illustrated in Figure 1 and outlined in the following steps:

\begin{enumerate}
 \item[\textbf{$a$}.] In a WD-NS system, the WD, with a mass ratio of $q>q_{\rm crit}$ is poised to commence the transfer of material when its orbit falls within the tidal radius (Figure 1(a)).
 \item[\textbf{$b$}.] The WD is disrupted by the NS, and the system undergoes a UMT. The mass-radius relation of massive WD can be approximately expressed as \citep{Nau1972,Mar2016}
 \begin{eqnarray}
 R_{\rm WD}&\approx& 1.0\times 10^{9}~{\rm cm}~\left(\frac{M_{\rm WD}}{0.7~M_{\odot}}\right)^{-1/3} 
 \nonumber\\ 
 &\times& \left[1-\left(\frac{M_{\rm WD}}{M_{\rm ch}}\right)^{4/3}\right]^{1/2},
 \end{eqnarray}
 where $M_{\rm ch}$ is the Chandrasekhar mass, and we take it as $1.45~M_{\odot}$. We presume a $1.4~ M_{\odot}\times 1.4~ M_{\odot}$ WD-NS binary system. The radius of the WD is $1.7\times 10^{8}\rm~cm$. This result is associated with that in \citet{Alt2005}. The radius of Roche lobe is \citep{Egg1983}
  \begin{eqnarray}
 R_{\rm ro}&=& \frac{0.49q^{2/3}}{0.6q^{2/3}+\ln \left(1+q^{1/3}\right)}a,
 \end{eqnarray}
 where $a$ is the separation of the binary, $q=M_{\rm WD}/M_{\rm NS}$ is the mass ratio of the binary. The UMT starts when $R_{\rm WD} = R_{\rm ro}$ implying that  $a\approx 4.5\times 10^{8}~\rm cm$. Then, the orbital period is $P=2\pi a^{3/2}[G(M_{\rm NS}+M_{\rm WD})]^{-1/2}\approx 3.1\rm~s$. This phase lasts for 3-10 orbital periods \citep{Pas2011, Bob2022}. During the disruption stage, the elongated WD interacts with itself, producing shocks and heating the material \citep{Zen2020, Bob2022}. The temperature and density are extremely high at the inner region of the accreting disk, which gives rise to nuclear burning. The simulation results from \citet{Bob2022} indicate that a very tiny fraction of material ($\sim $1\% of initial WD mass) is dynamically ejected at this stage, whereas their previous simulations \citep{Bob2017}, as well as those by \citet{Pas2011}, did not show any material being dynamically ejected during this phase (Figure 1(b)).
\end{enumerate}

In our analysis, we bifurcate the subsequent processes into two distinct scenarios. In the first scenario, the NS merges into the WD at the post-UMT stage,  resulting in a significant gravitational perturbation. This perturbation causes the immediate collapse of the TZlO. In the second scenario, the TZlO formed post-UMT does not undergo an immediate collapse. Instead, it remains stable for a finite cooling time before collapse. The time delay of the collapse implies a divergent evolutionary trajectory. Then, we delve into a detailed discussion of both scenarios. 

For the first scenario:

\begin{enumerate}
 \item[\textbf{$c1$}.] Most of the material tidally disrupted remains bound during the UMT process \citep{Pas2011, Zen2020, Bob2022, Kal2023}. A portion of the WD debris eventually envelops the NS, while the remainder forms a disk orbiting around the remnant \citep{Pas2011}. For the case of $q \approx 1$ that we are discussing, the NS may not disrupt the WD entirely during the UMT phase and will merge into the WD, resulting in a remnant surrounded by a disk \citep{Yang2022} (Figure 1~(c1)). The remnant is composed of WD material wrapped in an NS, and we call it TZlO. Numerical simulations \citep{Pas2011} have shown that the newborn disk, formed by the WD debris, is about the same size as the Roche limit, with its mass exceeding $50\%$ of initial WD mass. The TZlO is spherical on account of the rotation and with a little larger scale than the initial WD \citep{Pas2011}. Disregarding the minor portion of material dynamically ejected, the mass of the TZlO consists of the mass of NS plus a small fraction \citep[e.g., $\sim 30\% - 50\%$;][]{Pas2011} of the initial WD mass. If the hypothesis presented by \citet{Yang2022} in which the WD is not totally disrupted holds true, then the mass of the TZlO would be correspondingly larger. We propose the resultant mass of the TZlO is $\sim 2.0~M_{\odot}$. The abrupt impact of gravitational disturbance overcomes the support of electron degeneracy pressure, thermal pressure, and centrifugal force, inducing the envelope's collapse.

Although WDs are also compact objects, the merger process between a WD and an NS fundamentally differs from that of NS-NS/BH mergers. It more closely resembles the collapse process of a massive star, yet this collapse is distinct from the accretion-induced collapse \citep{Yi1998} associated with a WD accreting material from a companion star. Here, we describe this process as gravitational disturbance-induced collapse (GDIC). The freefall timescale of the GDIC can be roughly written as
\begin{eqnarray}
t_{\rm ff,TZ}&\sim& \left(\frac{R_{\rm TZ}^{3}}{G M_{\rm TZ}}\right)^{1/2} \nonumber\\
&=& 0.17~{\rm s}~\left(\frac{R_{\rm TZ}}{2\times 10^{8}~\rm cm}\right)^{3/2}\left(\frac{M_{\rm TZ}}{2.0~M_{\odot}}\right)^{-1/2}, \nonumber\\
\end{eqnarray}
 where $G$ is the gravitational constant, $R_{\rm TZ}$ is the radius of TZlO. The viscous timescale of the disk is 
\begin{eqnarray}
t_{\rm vis} &\simeq& 16.6~{\rm s}~\left(\frac{\alpha}{0.05}\right)^{-1}\left(\frac{H/R_{\rm disk}}{0.5}\right)^{-2}  \nonumber\\ 
&\times& \left(\frac{R_{\rm disk}}{4.5\times 10^{8}~\rm cm}\right)^{3/2}\left(\frac{M_{\rm TZ}}{2.0~M_{\odot}}\right)^{-1/2}, 
\end{eqnarray}
 where $\alpha$ is the viscosity constant, $H$ is the disk thickness, $R_{\rm disk}$ is the radius of the disk, and $M_{\rm TZ}$ is the mass of TZlO. 
 
 During the collapse process of the TZlO, the envelope WD material undergoes neutronization, and the forming proto-NS is surrounded by a neutron-rich disk that ejects a portion of neutron-rich material. This ejected material, combined with the dynamical ejecta from the process shown in Figure 1(b), moves as a unified entity \citep{Yang2022}. In Figures 1(d11) and (d12), we use different colors to represent these two distinct sources of ejecta.

  \item[\textbf{$d11$}.] If the remnant does not promptly collapse to a BH, the nascent NS, influenced by differential rotation, would generate subsequent high-energy emissions via the magnetic bubble mechanism \citep{Klu1998, Yang2022}. In this scenario, the GRBs generated are likely composed of two components: the jet produced by the accretion of the disk onto the NS \citep{Zhang2009, Zhang2010}, and the magnetic bubbles (the helical lines) from the differential rotation of the protomagnetar reveal as EE. The magnetic dipole radiation (the red-wavy lines with arrows) also contributes a certain amount of energy to the ejecta such that an optical excess will be observed during the afterglow phase.
 
 \item[\textbf{$d12$}.] Otherwise, the post-merger object collapses to a BH immediately. In comparison to the scenario of d11, the GRBs produced in this case consist solely of jets generated by the disk accreted onto the BH, without the part produced by magnetic bubbles. Consequently, the observed GRB would be classified as an ordinary long-duration burst devoid of any EE components. Furthermore, due to the absence of energy supply from magnetic dipole radiation, the optical excess in the ejecta would not be as pronounced as observed in the scenario of d11.

\end{enumerate}

For the second scenario:

\begin{enumerate}
 \item[\textbf{$c2$}.] The object formed post-merger, which has a mass distribution similar to that in case c1, possesses sufficient outward forces, including electron degeneracy pressure, thermal pressure, and centrifugal force, to balance the effects of its self-gravity. Consequently, the UMT results in a TZlO with a disk that will persist for a certain duration\footnote{In fact, dynamical ejecta similar to those in Figure 1 (c1) are present in this scenario as well. However, in this case, such ejecta do not produce significant observational effects, and hence, they are not marked in the figure.} (Figure 1(c2)). Because of the shock heating during the tidal disruption process, the envelope of the newly formed TZlO, as well as the disk, is very hot, with temperatures up to $1.5\times 10^{9}~\rm K$ \citep{Pas2011}. The numerical simulation conducted by \citet{Bob2022} unveiled elevated thermal conditions. Specifically, the disk's midplane is characterized by an assemblage of free nucleons, exhibiting both high-density and temperature conditions that are conducive to the nucleosynthesis of heavy elements, as delineated by \citet{Mar2016}. Concurrently, the outflow is accompanied by the formation and evolution of the disk.

\clearpage
\begin{figure}
\vspace{-0.1cm}
\hspace{-3.0cm}
\label{fig:1}
\includegraphics[width=1.5\textwidth]{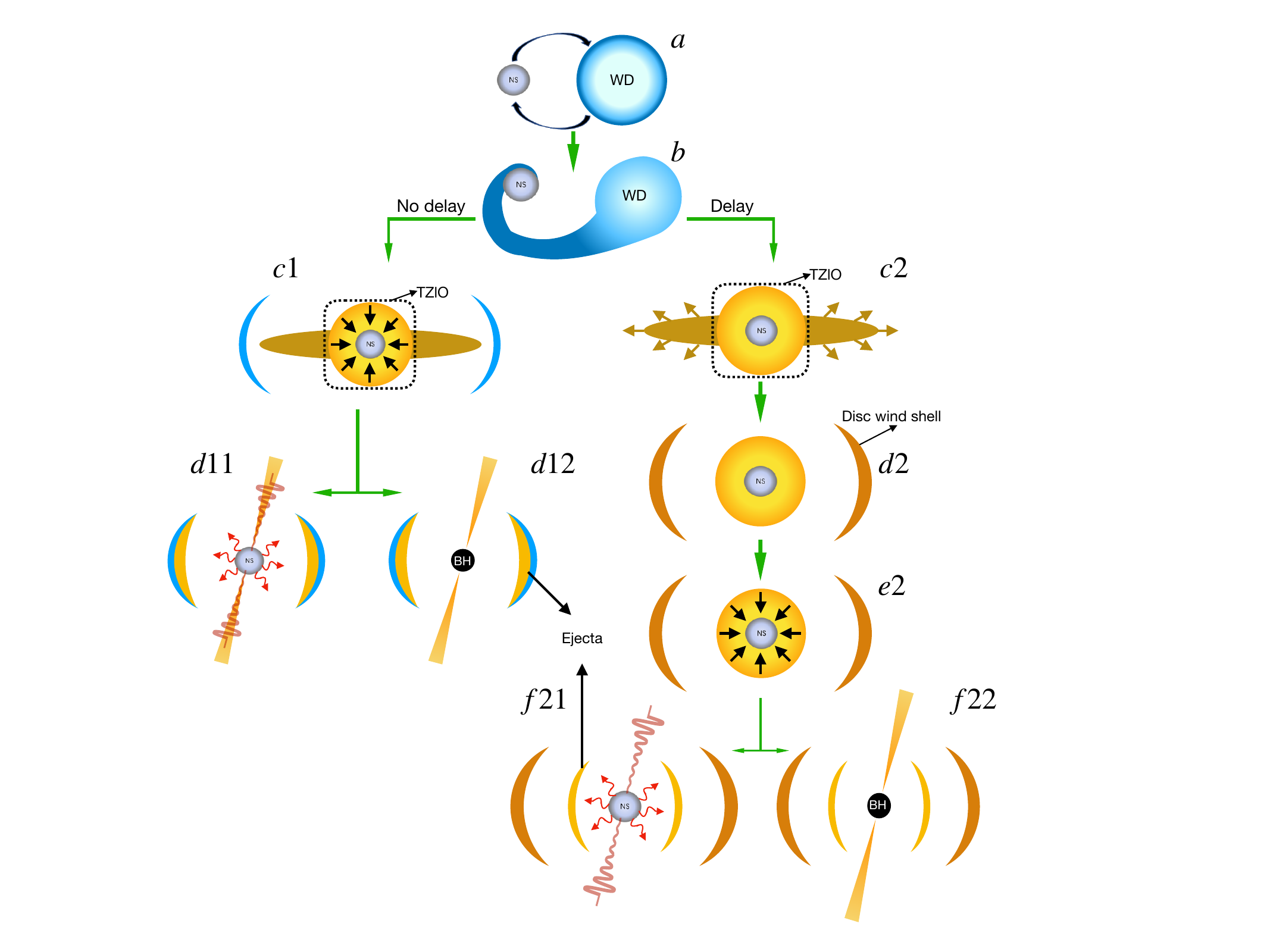}
\caption{The formation and collapse of a TZlO from a WD-NS merger. The post-merger outcomes based on whether the product immediately collapses into two scenarios: on the left, the formed TZlO collapses immediately, while on the right, the TZlO undergoes a period of cooling before collapsing. For both scenarios, the possibility of the collapse resulting in either a massive NS or a BH is discussed. In the cases where the product is an NS, we consider the magnetic dipole radiation (represented by red-wavy lines with arrows) and the radiation from magnetic bubbles (represented by the helical lines). }
\end{figure}
\clearpage

 \item[\textbf{$d2$}.] During the outflow process, part of the material will be lost in the form of disk wind, forming a shell, as shown in Figure 1~(d2). However, there is currently no consensus on how much material is ejected. The results from \citet{Zen2019} suggest that only a small fraction of the material is ejected, with the majority still bound by the NS. \citet{Met2012} and \citet{Fer2019} indicated that more than $50\%$ of the disk mass is lost. Although \citet{Bob2022} did not specify the exact amount of the mass of disk wind, they propose that more than $60\%$ of the disk mass would be ejected. \citet{Kal2023} believed that the outflow is greatly influenced by the entropy of the disk. If the disk is in a state of high entropy, the proportion of material lost could range from $\sim 20\%-40\%$ of the initial WD mass, while in a state of low entropy, this proportion could be smaller than $20\%$.
  
Several tens to thousands of seconds after the merger, the mass of the outflow ceases to increase significantly, indicating that the disk wind phase has essentially concluded \citep{Fer2019, Zen2019}. Eliminating the mass loss caused by disk wind, the remaining WD material and the NS together constitute the TZlO. The radius of the TZlO is close to the initial radius of WD \citep{Pas2011}. The mean density of the TZlO envelope is $\rho \sim 1.0\times 10^{8}~\rm g~cm^{-3}$. For the extreme relativistic condition, the calculated thermal gas pressure to electron degeneracy pressure of the envelope is $P_{\rm th}/P_{\rm e,ER}\approx 8.04\times 10^{-16}n k_{\rm B}T(\rho/\mu_{e})^{-4/3}\approx 0.61$, where $n$ is the particle number density. We take $T = 3\times 10^{9}~\rm K$ here, which is higher than the results presented by \citet{Pas2011} because of the more massive WD in our model. The high-temperature and high-density conditions have reached the critical condition for carbon burning. \citet{Zen2020} calculated the energy released by thermonuclear burning for $\sim 10^{46}-10^{47}~$erg. We noticed that the WD material at the innermost region was accreted by the NS continuously. The accretion luminosity could be represented by the Eddington luminosity. The thermal energy is $E_{\rm th} = \frac{16}{3}\pi R_{\rm TZ}^{3} \sigma T^{4}/c \sim 10^{49}\rm~ erg $. This indicates that nuclear energy has little contribution to the total energy. During this period, there is not significant mass outflow, the neutrino emission is the dominant cooling mechanism as the system is in high-temperature and high-density conditions \citep{Bea1967}. The cooling timescale can be calculated as $\tau_{\rm cooling}\approx 1.76\rho_{6}/T_{9}^{8}~\rm yr$ \citep{Pas2011}. When the temperature is set to be $3\times 10^{9}\rm~K$, the cooling timescale is approximately 1 week.

\item[\textbf{$e2$}.] During the process of b, c2 and d2, the magnetic field of the NS would be amplified as accreting materials from the WD \citep{Zho2020}. After cooling down, the TZlO eventually begins to collapse (Figure 1~(e2)) when the supporting force cannot resist the gravitational force. Such GDIC can naturally lead to a GRB central engine that consists of a central compact object, similar to the case of a long-GRB central engine, yet with much denser WD accreted material.

\item[\textbf{$f21$}.] During the collapsing of TZlO, the material of the envelope is neutronized. Part of the neutronized material would be ejected \citep{Yang2022}. The fate of the remnant is similar to d11. If the remnant does not collapse to BH promptly, the ejected material can be further heated by the magnetic dipole radiation of the central magnetar and radiate as an optical bump in observation. Furthermore, the ejected material can also interact with the disk wind shell at a larger radius and produce an additional late-time bump in the optical band.

\item[\textbf{$f22$}.] Otherwise, the remnant rapidly collapses to a BH (although such a case is unlikely, it is not entirely ruled out), and the ejecta cannot be accelerated. The ejecta requires a long time to catch up with the disk wind shell, and due to the ejecta not achieving subrelativistic speed, the collision energy is relatively low, resulting in no significant observable effects.

\end{enumerate}

\section{Model of the Afterglow}

The remnant of TZlO collapse can exist in either one of two states: a massive magnetar (Figure~1~(d11), (f21)) and direct collapse into a BH (Figure~1~(d12), (f22)). In the scenario where a BH is formed, the optical afterglow may exhibit an excess due to the decay of heavy elements, potentially produced by the $r$-process within the ejecta, in addition to the standard afterglow. In the scenario where a massive magnetar is formed, the magnetic dipole radiation significantly heats and accelerates the ejecta, leading to more pronounced observational effects in the afterglow. This may include features analogous to those observed in a merger-nova \citep{Yu2013}, as well as components characteristic of an interaction-powered supernova (IPSN). Here, we consider the magnetar scenario, and the observed flux of such emissions can be derived as follows. 

The total rotation energy of the newly born twirling magnetar is 
\begin{equation}
E_{\rm rot}=\frac{1}{2} I \Omega^{2},
\end{equation}
where $I$ is the moment of inertia, and $\Omega$ is angular velocity. The energy loss follows
\begin{eqnarray} \label{eq:Erot}
-\frac{dE_{\rm rot}}{dt}=-I\Omega \dot{\Omega}&=&\frac{B_{\rm p}^{2}R^{6}\Omega^{4}}{6c^{3}}+\frac{32GI^{2}\epsilon^{2}\Omega^{6}}{5c^{5}} \nonumber\\
&=&L_{\rm EM}+L_{\rm GW},
\end{eqnarray}
which incorporates electromagnetic luminosity, $L_{\rm EM}$, and gravitational wave (GW) luminosity, $L_{\rm GW}$. $B_{\rm p}$ corresponds to the NS surface magnetic field at the pole. $\epsilon$ is the ellipticity of the magnetar. The time derivative of angular velocity can be described as
\begin{equation} \label{eq:om}
\dot{\Omega}=-k\Omega^{n},
\end{equation}
where $k$ can be simplified as a constant, and $n$ is the braking index \citep{Las2017}. Given that the deformation of the protomagnetar is minimal following the cessation of differential rotation, the contribution of GW radiation can be considered negligible. Following \citet{Yu2013} and \citet{Las2017}, we set $n=3$ in this work.

Combining electromagnetic part of Eq.(\ref{eq:Erot}) with Eq.(\ref{eq:om}), one can easily derive $L_{\rm EM}$ in the simple form of
\begin{equation}
L_{\rm EM} = L_{0}\left(1+\frac{t}{\tau}\right)^{-2},
\end{equation}
where $L_{0}=10^{49}R_{\rm s,6}^{6}B_{\rm p,15}^{2}P_{0,-3}^{-4}\rm erg~s^{-1}$ is the initial luminosity of the magnetar, and $\tau=2\times 10^{3}R_{\rm s,6}^{-6}B_{\rm p,15}^{-2}P_{0,-3}^{2}~\rm s$ is the spindown timescale of the magnetar. So, $L_{\rm EM}$ can be regarded as the spindown luminosity of the magnetar, $L_{\rm EM}=L_{\rm sd}$.

The total energy of the ejecta can be expressed as
\begin{equation}
E_{\rm ej}=\Gamma M_{\rm ej}c^{2}+\Gamma E_{\rm int}^{\rm RF}+\Gamma^{2}M_{\rm sw}c^{2},
\end{equation}
where $\Gamma$ is the Lorentz factor, $E_{\rm int}^{\rm RF}$ is the internal energy in the comoving rest frame, $M_{\rm sw}=4/3\pi R^{3}n_{0}m_{\rm p}$ is the mass of the swept-up circumburst medium (CBM), and $R$ is the radius of the ejecta. 

Assuming a fraction $\xi$ of spindown energy injected into the ejecta, one can calculate that the energy change of the eject is
\begin{equation}
dE_{\rm ej}=(\xi L_{\rm sd}-L_{\rm e})dt,
\end{equation}
 where $L_{\rm e}$ is the radiated bolometric luminosity. 
 
The dynamical evolution of the ejecta can be written as \citep{Hua2000,Yu2013}
\begin{equation}
\frac{d\Gamma}{dt} = \frac{\xi L_{\rm sd}-L_{\rm e}-\Gamma D \frac{dE_{\rm int}^{\rm RF}}{dt^{\rm RF}}-(\Gamma^{2}-1)c^{2}\frac{dM_{\rm sw}}{dt}}{M_{\rm ej}c^{2}+E_{\rm int}^{\rm RF}+2\Gamma M_{\rm sw}c^{2}},
\end{equation}
where $D = 1/[\Gamma(1-\beta)]$ is the Doppler factor.

The variation of internal energy of the ejecta in the comoving frame can be written as \citep{Kas2010}
\begin{equation} \label{eq:rad}
\frac{dE_{\rm int}^{\rm RF}}{dt^{\rm RF}} = \xi \frac{L_{\rm sd}}{D^{2}}-L_{\rm e}^{\rm RF}-P^{\rm RF}\frac{dV^{\rm RF}}{dt^{\rm RF}},
\end{equation}
where the comoving radiated luminosity is $L_{\rm e}^{\rm RF}=L_{\rm e}/D^{2}$. 

The last term of Eq.(\ref{eq:rad}) represents the work of free expansion. In the comoving frame of the ejecta, the pressure is dominated by radiation, so 
\begin{equation}
P^{\rm RF}=E_{\rm int}^{\rm RF}/3V^{\rm RF},
\end{equation}
and the evolution of the bulk of ejecta is
\begin{equation} 
\frac{dV^{\rm RF}}{dt^{\rm RF}}=4\pi R^{2}\beta c.
\end{equation}

By utilizing
\begin{equation} 
\frac{dR}{dt}=\frac{\beta c}{1-\beta},
\end{equation}
one can write the comoving radiation bolometric luminosity as \citep{Kas2010,Kot2013}
\begin{equation}
L_{\rm e}^{\rm RF}=\begin{cases}
\frac{E_{\rm int}^{\rm RF}\Gamma c}{\tau R}, & \tau > 1 \\
\frac{E_{\rm int}^{\rm RF}\Gamma c}{R}, & \tau < 1, \\
\end{cases}
\end{equation}
where $\tau=\kappa (M_{\rm ej}/V^{\rm RF})(R/\Gamma)$ is the optical depth of the ejecta, and $\kappa$ is the opacity. 

The dynamical evolution of the ejecta properties, such as $\tau$, $R$, $E_{\rm int}^{\rm RF}$, and $V^{\rm RF}$ can be solved through Eqs. (8)-(15). 

For any given frequency, $\nu$, the observed specific flux of the ejecta can be calculated as
\begin{equation} 
F_{\rm \nu,ej}=\frac{1}{\rm max(\tau,1)}\frac{1}{4\pi D_{\rm L}^{2}}\frac{8\pi^{2}D^{2}R^{2}}{h^{3}c^{2}\nu}\frac{(h\nu/D)^{4}}{{\rm exp}(h\nu/D k T^{\rm RF})-1},
\end{equation}
where $D_{\rm L}$ is the luminosity distance of the burst, $h$ is the Planck constant, $k$ is the Boltzmann constant, and $T^{\rm RF}=(E_{\rm int}^{\rm RF}/a V^{\rm RF})^{1/4}$ is the comoving temperature.

In addition, the interaction between ejecta and the disk wind shell at a large radius (Figure 1 (f21)) causes a forward shock and reverse shock. Such shocks heat the shell and lead to an IPSN in the optical. Following \citet{Cha2012} and the parameterization therein, the output luminosity of the interaction can be expressed semi-analytically as\footnote{As our methodology closely adheres to the procedures outlined by \citet{Cha2012}, readers interested in this aspect are encouraged to consult that paper for a detailed exposition.}
\begin{eqnarray}
L_{\rm t,IPSN}&=&\frac{1}{t_{0}}e^{-\frac{t}{t_{0}}}\int^{t}_{0}e^{\frac{t'}{t_{0}}}[\frac{2\pi}{(n-s)^{3}}g^{n\frac{5-s}{n-s}}q^{\frac{n-5}{n-s}}(n-3)^2 \nonumber \\
&\times& (n-5) \beta_{\rm F}^{5-s}A^{\frac{5-s}{n-s}}(t'+t_{i})^{\frac{2n+6s-ns-15}{n-s}} \nonumber \\
&\times& \theta (t_{\rm FS,BO}-t')+2\pi \left(\frac{Ag^{n}}{q}\right)^{\frac{5-n}{n-s}}
\beta_{\rm R}^{5-n}g^{n}\nonumber\\
&\times& \left(\frac{3-s}{n-s}\right)^{3} (t'+t_{i})^{\frac{2n+6s-ns-15}{n-s}}\theta (t_{\rm RS,*}-t')]dt'. \nonumber \\
\end{eqnarray}
The final observed flux can be calculated by combining Eqs. (16) and (17):
\begin{equation} 
F(t, \nu,P) =F_{\rm \nu,AG}(t)+F_{\rm \nu,ej}(t,t_{0,1})+\frac{L_{\rm t,IPSN}(t,t_{0,2})}{4\pi D_L^2}
\end{equation}
where $P$ is set free parameters as specified in \S 4.3, $t_{0,1}$ and $t_{0,2}$ are the onset time of the two components. $F_{\rm \nu,AG}$, includes the standard afterglow component using an external forward shock model following the Python $afterglowpy$ module \citep{Rya2020}. Eq. (19) can be directly compared with observational data through a Monte Carlo fit (\S 4.3). 

\section{Explanations for two peculiar GRBs}

\subsection{GRB 211211A}

The prompt emission of GRB 211211A exhibits a $T_{90}$ of 43.18 s, which can be decomposed into two distinct phases: a main emission (ME) lasting 13 s, followed by EE persisting for 55 s \citep{Ras2022, Tro2022, Yang2022}. Subsequent observations have identified a host galaxy for the burst, located at a redshift of $z=0.076$. The characteristics of GRB 211211A align with those typically observed in Type I GRB, including \citep{Yang2022} (1) minimal spectral lag; (2) the burst has a large offset within its host galaxy, which is consistent with the expected progenitor sites for Type I GRB; (3) the ME comply with the Amati relation for Type I GRB (whereas the EE follows to Type II GRB in the Amati relation diagram); (4) optical excess around 1-10 days post-burst; and (5) the absence of an associated SN component. However, the duration of the prompt emission is characteristic of Type II GRB, which is not typically thought to be produced by binary NS mergers due to the difficulty in sustaining such prolonged emission.

\citet{Yang2022} postulated that GRB 211211A might originate from a WD-NS merger. They suggested that when the mass of WD approaches the Chandrasekhar limit, the NS merger into the WD triggers the remnant to collapse under intense gravitational forces. This collapse is posited to generate the ME of GRB 211211A. The protomagnetar, characterized by its differential rotation, is theorized to produce magnetic bubbles \citep{Klu1998}, which correspond to the observed EE component. The collapse of the TZlO envelope material, being a neutronization process, results in some neutrons incorporated into the ejecta, facilitating the $r$-process within this material \citep{Whe1998, Yang2022}. Additionally, the nascent magnetar is conjectured to provide energy to the ejecta \citep{Ai2022}, contributing to the optical excess observed from 1-10 days post-burst \citep{Yang2022}. This scenario provides a comprehensive explanation for the multifaceted emission profile of GRB 211211A, encompassing both the prompt emission and its subsequent afterglow features. This corresponds to the first scenario in which the TZlO collapses with no delay in our model.

In the literature, there have been attempts to reconcile this burst within the framework of binary NS mergers. \citet{Gao2022} propose that a post-merger compact object with a strong magnetic flux can extend the accretion timescale, thereby enabling binary NS mergers to produce longer-duration bursts. \citet{Got2023} suggest that a binary merger with a significant mass ratio can result in a substantial accretion disk ($M_{\rm d}>$ 0.1 $M_{\odot}$), which extends the accretion timescale and could account for the observed duration of the burst. Furthermore, they posit that when the accretion disk enters a magnetically arrested disk phase, the accretion rate decreases $t^{-2}$ over time, which could explain the genesis of the EE. \citet{Zhong2023} posited that in the case of a WD-NS merger, if the resulting accretion disk is in low entropy and efficient wind, the accretion onto the NS can meet requirements of a GRB \citep{Kal2023}. The prompt emission is attributed to magnetic bubbles generated by the differential rotation of the NS induced by accretion.

\subsection{GRB 200826A}

GRB 200826A was observed to have a duration of approximately 1 s. In a detailed analysis of this burst, \citet{Zha2021} contended that the influence of the ``tip-of-the-iceberg" effect is absent, suggesting that the 1 s duration represents the intrinsic timescale of the burst, aligning with a Type I GRB. However, when positioned on the Amati relation, the burst unexpectedly falls within the Type II GRB region.  Optical observations of the burst revealed several excess emission points occurring $\sim 20$ days post-burst. \citet{Ros2022} argue that these excess points bear a resemblance to SN 1998bw and likely represent an SN component associated with the burst. The occurrence of such a short-duration GRB, presumably originating from the collapse of a massive star, remains puzzling. While there have been previous observations of Type II GRBs with durations shorter than 2 s, such as GRB 090426, these were shown to be significantly influenced by the tip-of-the-iceberg effect, with intrinsic durations far exceeding 2 s \citep{Lv2014}. \citet{Ahu2021} hypothesize that GRB 200826A represents a rare instance where the burst managed to penetrate the massive stellar envelope in the final 1 s. \citet{Wang2022} proposed an alternative explanation, suggesting that the burst could be a precursor of a long GRB similar to GRB 160625B, with the main burst phase being obscured due to jet precession or obstruction by a companion star.

\citet{Zha2021} noted that the progenitor of GRB 200826A could be a WD-NS binary system. This system underwent UMT events prior to the merger, characterized by intense material interactions that significantly elevated its temperature. Following the merger of the WD and NS, a thermally supported compact object exceeding the Chandrasekhar limit, referred to as a TZlO, was formed. This object temporarily resisted gravitational collapse due to the combined effects of thermal pressure, rapid rotation, and electron degeneracy pressure.

The accretion disk formed during the merger rapidly lost a substantial amount of material due to disk wind, leading to the formation of an expanding outer shell. As neutrino radiation continued, the TZlO shed most of its thermal energy, eventually succumbing to internal gravitational forces and collapsing. This collapse process is accompanied by neutronization, resulting in the ejection of neutron-rich material \citep{Yang2022}.

The protomagnetar exhibits significant differential rotation during this process, generating magnetic bubbles that were the source of the prompt emission of GRB 200826A. Subsequently, the magnetar's dipole radiation accelerated the ejecta, producing radiation characteristics akin to those of a merger-nova \citep{Yu2013}. This material accelerated to subrelativistic velocity and eventually caught up with the earlier-formed shell by disk wind, colliding with it and triggering SN-like radiation.

\subsection{The Fit}

\begin{figure}
\hspace{-0.1cm}
\includegraphics[width=0.5\textwidth]{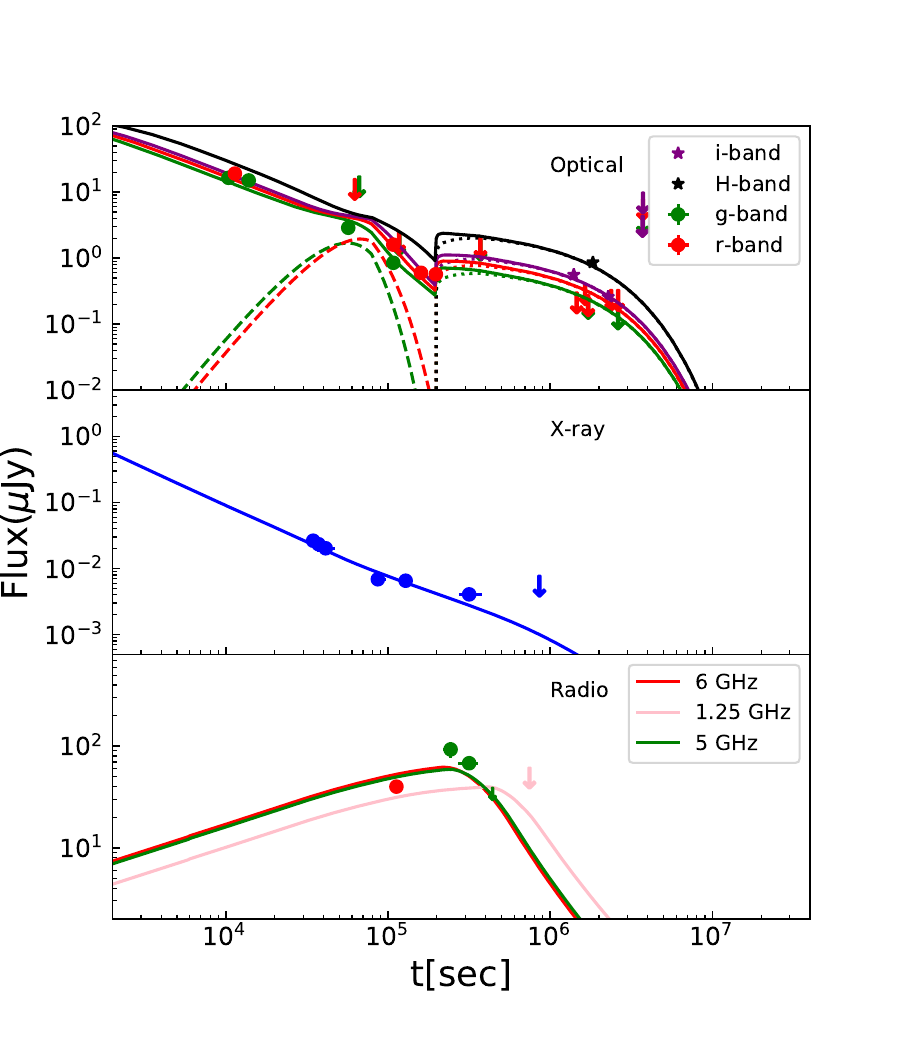}

\caption{Multiwavelength observation data of GRB 200826A overplotted with our model using the best-fit parameters. \textit{Top:} data points show the optical observations. Solid lines show the model predication using the best-fit parameters in Table \ref{tab:parameters}. The dashed lines show the merger-nova-like components and the dotted lines show the late-time SN-bump-like component. \textit{Middle:} data points show the X-ray observations by Swift/XRT. The solid line shows the model prediction. \textit{Bottom:} radio afterglow (filled circles) overplotted with the best-fit models (solid lines). }
\label{fig:2}
\end{figure}

Given that the afterglow of GRB 211211A has been effectively modeled by \citet{Yang2022}, this study refrains from replicating that analysis and instead focuses on further modeling the afterglow of GRB 200826A. To do so, we first collect all available observational data of GRB 200826A in the following energy bands:
\begin{itemize}
 \item $\gamma$-ray. Following \citet{Zha2021}, GRB 200826A is shown as a short-duration burst with $T_{90} = 0.96^{+0.05}_{-0.08}\rm~s$ at $10-800~\rm keV$ band, with an isotropic energy of $E_{\rm iso} = (7.09\pm 0.28)\times 10^{51}\rm erg$.
 
 \item X-ray. Swift/X-Ray Telescope (XRT) skewed in the direction of GRB 200826A at $\sim 6\times 10^{4}~\rm s$ after the trigger time of Fermi/Gamma-ray Burst Monitor (GBM). 
 The X-ray data are obtained from the Swift/XRT repository\footnote{\url{https://www.swift.ac.uk/xrt_curves/00021028/}} \citep{Eva2007,Eva2009} and are plotted in the middle panel of Figure \ref{fig:2}.

 \item Optical. The first optical counterpart is obtained by the Zwicky Transient Facility (ZTF) at $\sim 0.2$~days \citep{Ahu2020b}, and the host galaxy is confirmed with redshift at $z=0.7481\pm 0.0003$ \citep{Rot2020}. An optical flare is observed by the Large Binocular Telescope \citep[LBT;][]{Rot2020} at 2.18 days. We carried out our observations in the $r$-band with the Las Cumbres Observatory Global Telescope (LCOGT) and obtained several upper limits around 2.27 days. \citet{Ahu2021} reported a series of $i$-band observations from the Gemini-North 8 m telescope and measured an optical bump with $M_i= 25.45\pm 0.25\rm~AB\rm~mag$ at $t$= 28.28 days. \citet{Ros2022} reprocessed the observation and revised it to $M_i= 24.53\pm 0.21\rm~AB\rm~mag$, and they found the other two optical observations. One was observed by LBT at $t$= 37.1 days after the burst with $M_H= 24.06\pm 0.20\rm~AB\rm~mag$. The other is at $t$= 46.11 days, and the magnitude was $M_i= 25.36\pm 0.26\rm~AB\rm~mag$. All those data, as well as those reported in GCN circulars\footnote{\url{https://gcn.gsfc.nasa.gov/other/200826A.gcn3}} by multiple facilities, including the Telescopio Nazionale Galileo (TNG), Gran Telescopio Canarias (GTC), and Lowell Discovery Telescope (LDT), are all collected and listed in Table \ref{tab:optical}.

\clearpage
\begin{longtable}{c c c c c c c} 
\caption{Observations of optical counterpart of GRB~200826A.}\\
\hline 
$\delta t$(day) & Telescope & Band & System & Magnitude & Flux Density ($\rm \mu$Jy) & Ref.\\
\hline
\multicolumn{7}{c}{Optical Counterpart}\\
\hline
0.21 & ZTF & $g$ & AB & 20.86$\pm 0.04$ & 16.44$_{-0.59}^{+0.62}$ & (1)\\[4pt]
0.23& ZTF & $r$ & AB & 20.70$\pm 0.05$ & 19.05$_{-0.86}^{+0.90}$ & (1) \\[4pt]
0.28 & ZTF & $g$ & AB & 20.96$\pm 0.16 $ & 15.00$_{-2.05}^{+2.38}$ & (1)\\[4pt]
0.77 & Kitab-ISON RC-36 & CR & AB & $>$ 20.2 & $<$ 30.30 & (3)\\[4pt]
1.15 & ZTF & $g$ & AB & 22.75$\pm 0.26$ & 2.88$_{-0.61}^{+0.78}$ & (1)\\[4pt]
1.21 & ZTF & $r$ & AB & $>$ 21.30 & $<$ 10.96 & (1)\\[4pt]
1.29 & ZTF & $g$ & AB & $>$ 21.20 & $<$ 12.02 & (1)\\[4pt]
1.74 & Kitab-ISON RC-36 & CR & AB & $>$ 20.4 & $<$ 25.12 & (3)\\[4pt]
1.79 & Swift/UVOT & White/(170-650 nm) & AB & 21.86$\pm 0.13$ & 8.55$_{-0.74}^{+0.83}$ & (2)\\[4pt]
2.18 & LBT & $g$ & AB & 24.08$\pm 0.19$ & 0.85$ _{-0.14 }^{+0.16}$ & (5)\\[4pt]
2.18 & LBT & $r$ & AB & 23.39$\pm 0.16$ & 1.60$ _{-0.22 }^{+0.25}$ & (5)\\[4pt]
2.27 & LCOGT & $r$ & AB & $>$ 23.30 & $<$ 1.74 & (9)\\[4pt]
2.28 & LCOGT & $g$ & AB & $>$ 23.41 & $<$ 1.57 & (9)\\[4pt]
3.23 & LDT & $r$ & AB & 24.46$\pm 0.12$ & 0.60$_{-0.06}^{+0.07}$ & (4)\\[4pt]
3.99 & GTC & $r$ & AB &  24.51$\pm 0.14$ &  0.57$ _{-0.07 }^{+0.08}$ & (8)\\[4pt]
28.28 & Gemini-North 8 m telescope/GMOS-N & $i$ & AB & 24.53$\pm 0.21$ & 0.56$_{-0.10}^{+0.12}$ & (7) \\[4pt]
28.28 & Gemini-North 8 m telescope/GMOS-N & $r$ & AB & $>$ 25.6 & $<$ 0.21 & (7)\\[4pt]
31.9 & TNG telescope & $r$ & AB & $>$ 25.3 & $<$ 0.27 & (6)\\[4pt]
37.1 & LBT telescope /LUCI & $H$ & AB & 24.06$\pm 0.20$ & 0.86$ _{-0.15 }^{+0.17}$ & (6)\\[4pt]
46.11 & Gemini-North 8 m telescope/GMOS-N & $i$ & AB &  25.36$\pm 0.26$ & 0.26$ _{-0.06 }^{+0.07}$ & (7) \\[4pt]
46.11 & Gemini-North 8 m telescope/GMOS-N & $r$ & AB & $>$ 25.5 & $<$ 0.23 & (7)\\[4pt]
72.05& LCOGT network observatory & $i$ & AB & $>$ 22.7 & $<$ 3.02 & (9)\\[4pt]
72.48 & Lijiang 2.4 m telescope & $g$ & AB & $>$ {22.58} & $<$ 3.37 & (9)\\[4pt]
72.48 & Lijiang 2.4 m telescope & $r$ & AB & $>$ {22.07 } & $<$ 5.40 & (9)\\[4pt]
72.48 & Lijiang 2.4 m telescope & $i$ & AB & $>$ {21.81} & $<$ 6.85 & (9)\\[4pt]
\hline
 		
\label{tab:optical}
\end{longtable}
\begin{tablenotes}
 \item (1) GCN Circular 28295  \citep{Ahu2020b}
 \item (2) GCN Circular 28300  \citep{Dai2020}
 \item (3) GCN Circular 28306  \citep{Bel2020}
 \item (4) GCN Circular 28312  \citep{Dic2020}
 \item (5) GCN Circular 28319  \citep{Rot2020}
 \item (6) GCN Circular 28949  \citep{Ros2020}
 \item (7) GCN Circular 29029  \citep{Ahu2020a}, \citet{Ros2022}
 \item (8)  \citet{Ros2022}
 \item (9) This work

\end{tablenotes}
\clearpage

 \item Radio. As reported in \citet{Ale2020}, the Very Large Array observed the position of GRB~200826A 2.28 days after the GBM trigger at a mean frequency of 6 GHz. A radio source was detected with a flux density of $\sim 40\rm~\mu Jy$ at the position of GRB 200826A. \citet{Rho2021} carried out a survey of enhanced Multi-Element Remotely Linked Interferometer Network observations of GRB~200826A. They obtained two radio detections of $93\pm 16$ and $68\pm 8\rm~\mu Jy$ at 5 GHz at $4.92\pm 0.5$ and $6.4\pm 0.9$ days, respectively, and an upper limit of $F_{\nu}<34 ~\rm \mu Jy$ at $8.7\pm 1.9$ days. The upgraded Giant Metrewave Radio Telescope performed an observation 14.47 days after the burst. The central frequency is 1.25 GHz, and the radio emission reported was below a $3\sigma$ limit of 48.6$~\rm \mu Jy~beam^{-1}$. All those numbers are plotted in Figure 2 and taken into account in our fit.

\end{itemize}

We then fit our model (Eq. (19)) to the above observational data using a self-developed Bayesian Monte Carlo fitting package, McEasyFit \citep{Zha2015}, which ensures that the reliable best-fit parameters and their uncertainties can be realistically determined by the converged Monte Carlo chains. The free parameters, as well as their allowed ranges, are listed in Table \ref{tab:parameters}. The priors of the fitting free parameters are set to uniform (those listed as ``parameter name" in Table \ref{tab:parameters}) or log uniform (those listed as ``log parameter name" in Table \ref{tab:parameters}) distributions in physically allowed large ranges (see below). Our model successfully fit the data. The best-fit parameters, as well as their constraints, are listed in Table \ref{tab:parameters} and plotted in Figure \ref{fig:3}. The model predictions using the best-fit parameters are overplotted in Figure \ref{fig:2}.

Our fit highlights the following physical constraints of the system:
\begin{itemize}
 \item $\xi$. The best-fit value of the energy ejection ratio, $\xi$, is constrained at $0.35_{-0.09}^{+0.18}$, indicating that one-third of the spindown energy is injected into the ejecta. 
 
 \item $\kappa$. Our result shows the opacity of the ejecta is $\kappa = 19.40_{-4.73}^{+6.56}$. Such a large opacity suggests the ejecta should be rich in heavy elements \citep{Yu2018,Tan2020,Yang2022}, possibly brought by the effect of the neutron capture process.
 
 \item $M_{\rm ej}$. With $M_{\rm ej} =0.005_{-0.0017}^{+0.00008}M_{\odot}$, the mass of the ejecta during the collapse of the TZlO is found to be similar to the result of the NS-NS merger simulation \citep{Hot2013}.
 
 \item $n_{0}$. The interstellar medium density is contained at $11.80_{-3.03}^{+10.66} ~\rm cm^{-3}$. Such a large value is consistent with the observational fact that the burst is located in the host galaxy with a small offset \citep{Zha2021}, likely a denser area with a higher star-formation rate, yet it allows the possibility of forming a WD-NS system \citep{Too2018, Zho2020}. 

 \item $\theta_{\rm C}~\&~\theta_{\rm obs}$. The jet-core angle and the observer angle are constrained at $\theta_{\rm C}=12.32_{-1.55}^{+1.12} $ deg and $\theta_{\rm obs}=4.58_{-0.21}^{+3.95}$ deg, respectively, suggesting an on-axis observation of a typical GRB jet. 
 
 \item NS properties. Our fitting results constrain an NS with a magnetic field of 3.40$_{-0.72}^{+0.60} \times 10^{15}$ G and a rotating period of 2.63$_{-0.20}^{+5.59}$ ms, fully in agreement with the magnetar-powered GRB central engine. 

\begin{figure*}
\vspace{-1.0cm}
\hspace{-2.4cm}
\includegraphics[width=1.15\textwidth]{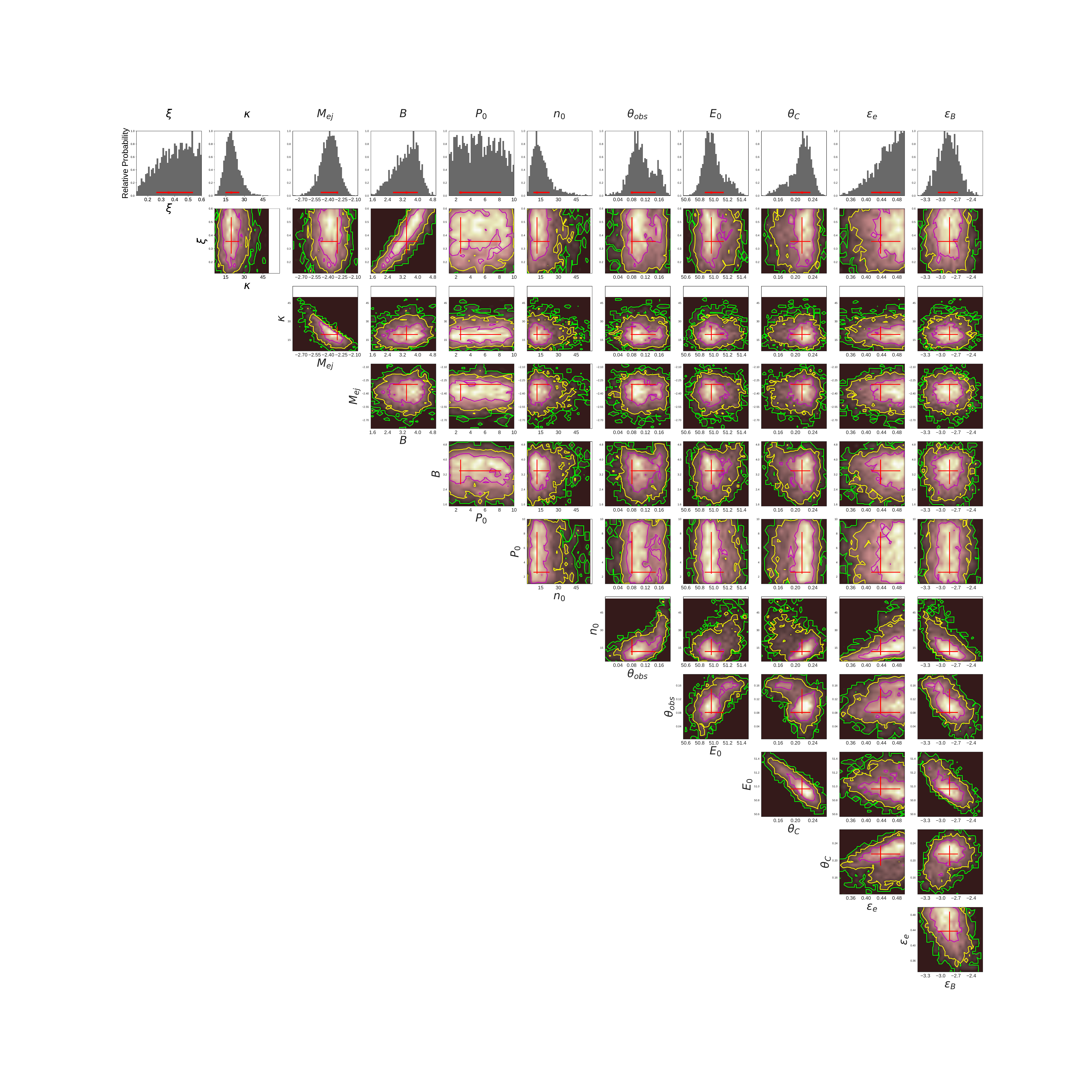}
\caption{Parameter constraints of our model fit using \textit{McEasyFit}. }
\label{fig:3}
\end{figure*}

 \item A merger-nova-like component. The best fit of our model yields a significant optical bump at $\sim 2-3$ days, as plotted in the top panel in Figure 2.
 
 \item The SN-like optical bump. Our model successfully explained the optical bump as an IPSN with the following key settings in Eq. (18): (1) compact profile $n = 5.1$. (2) The power-law exponent for the CBM shell density profile $s = 2$. (3) The mass of the disk wind shell is set to $M_{\rm S}= 0.5~M_{\odot}$. Assuming the maximum velocity of the disk wind to be $5\times 10^{8}\rm~cm~s^{-1}$ \citep{Zen2020, Bob2022}, we derive the mean density of the disk wind shell to be $\rho_{\rm S}=3\times 10^{-12}\rm g~cm^{-3}$ 10 days after the merger. (4) The radius when the ejecta collides with the disk wind shell is at $\sim3\times 10^{14}~\rm cm$. (5) The optical depth of the disk wind shell is $\tau_{\rm Thomson} \approx \kappa_{0} M_{\rm S}/4\pi R^{2} >1$, where $R$ is the radius of the disk wind shell and $\kappa_{0}$ is the Thomson scattering opacity of the disk wind shell, which is set to $\kappa_{0} = 0.4~\rm cm^{2}g^{-1}$. We notice that $R\lesssim 4.4\times 10^{15}~\rm cm$ meets the optically thick condition.

\end{itemize}

\begin{table}
\caption{Best-fit parameters and their uncertainties of our model to the multi-wavelength data.}
\begin{tabular}{ccc}
\hline 
Parameters & Range & Best Fit \\
\hline
$\xi$ & [0.1, 0.6]&$0.35_{-0.09}^{+0.18}$ \\[4pt]
$\kappa$&[0.1, 60] &$19.40_{-4.73}^{+6.56}$ \\[4pt]
$log(M_{\rm ej})$($M_{\odot}$) &[-5.0, -2.0] &$-2.30_{-0.18}^{+0.007}$ \\[4pt]
$B_{15}$ (G) &[0.1, 5.0] &$3.40_{-0.72}^{+0.60}$ \\[4pt]
$P_{0,-3}$ (s) & [1.0, 10]&$2.63_{-0.20}^{+5.59}$ \\[4pt]
$n_{0}$ & [0.1, 60]&$11.80_{-3.03}^{+10.66}$ \\[4pt]
$\theta_{\rm obs}$(rad) &[0.0, 0.2] &$0.08_{-0.004}^{+0.07}$ \\[4pt]
$\theta_{\rm C}$(rad) &[0.05, 0.4] &$0.22_{-0.03}^{+0.02}$ \\[4pt]
$log(E_{0})$(erg) & [50, 53]&$50.96_{-0.10}^{+0.18}$ \\[4pt]
$\epsilon_{\rm e}$ & [0.01, 0.5]&$0.44_{-0.02}^{+0.05}$ \\[4pt]
$\epsilon_{\rm B}$ &[-5.0, -2.0] &$-2.83_{-0.23}^{+0.16}$ \\[4pt]
\hline
 		
\label{tab:parameters}
\end{tabular}%
\end{table}

\subsection{Interpretation of the disparate properties of the two bursts}

Our model operates within the framework of a massive WD-NS merger, where delayed collapse leads to distinct properties of the two GRBs and follow-up observations.

\begin{itemize}

\item{In the first scenario, GRB 211211A is produced with a long-duration ME accompanied by EE, whereas GRB 200826A is likely a product of the second scenario, characterized by a short duration. For GRB 211211A, the characteristic timescale of the ME is decided by the viscous timescale of the accretion disk, while the EE results from the differential rotation of the post-collapse magnetar, which generates magnetic bubbles that burst forth. 

In the case of GRB 200826A, the delayed collapse allows for the evaporation of the accretion disk formed during the merger, with the GRB arising solely from the eruption of magnetic bubbles. The NS at the core of the TZlO continuously accretes envelope material prior to the collapse, thereby amplifying its magnetic field \citep{Zho2020}. This process results in a reduced timescale for magnetic bubble generation compared to that of GRB 211211A. If this process amplifies the surface magnetic field of the NS within the TZlO to $5\times 10^{14}\rm~G$, the interval time for the generation of magnetic bubbles would be approximately 0.03 s. Under the same differential kinetic energy ($E_{\rm d}$) and speed ($\Delta \Omega$) (e.g., $E_{\rm d}\approx 3\times 10^{51}\rm~erg$ and $\Delta \Omega \approx 10^{4}\rm~s^{-1}$ ) as assumed in \citet{Yang2022}, the duration of the magnetic bubbles would decrease to about 1 s, consistent with the prompt emission timescale observed in GRB 200826A.}

\item{Typically, long and short GRBs fall into their respective regions on the Amati relation diagram. However, plotting these two GRBs on the Amati relation diagram reveals some peculiarities. The ME of GRB 211211A is located at the site of Type I GRBs in the Amati relation, while the EE aligns with the site of Type II GRBs \citep[see Figure 2$b$ of][]{Yang2022}. GRB 200826A, although classified as a short burst, is positioned within the Type II GRBs category \citep[see Figure 2$a$ of][]{Zha2021}. This could be attributed to the similar origins of the EE of GRB 211211A and the prompt emission of GRB 200826A, which share comparable spectral properties.}

\item{No SN component was observed in association with GRB 211211A, whereas SN features were detected for GRB 200826A. This is because GRB 211211A triggered an immediate collapse following the WD-NS merger, while GRB 200826A experienced a delayed collapse after the merger. Before the collapse, a shell was formed by the disk wind on the outside (Figure 1~(d2)). Ultimately, cooling led to the collapse of the TZlO, and the ejecta, accelerated to subrelativistic speed, collided with the shell, producing optical radiation akin to that of an SN.}

\end{itemize}

\section{Summary and Discussion}

In this paper, we have introduced the collapse of TZlO as one of the progenitors of GRBs. In our discussion, there are two distinct scenarios of the collapse of a TZlO. In the first scenario, the TZlO undergoes immediate collapse post-formation. Due to the presence of a large-scale accretion disk generated during the merger process, the resulting GRB exhibits a long duration. Post-collapse, the object can evolve in two ways: if the TZlO collapses into a massive magnetar, it may produce a GRB similar to GRB 211211A. If the TZlO collapses directly into a BH, the resulting GRB will be a general long burst without an extended emission component.

In the second scenario, the TZlO, existing beyond the Chandrasekhar limit, is initially sustained by outward forces, allowing it to survive for a period. The accretion disk formed during the merger evaporates in the form of disk winds, creating an outward-moving shell composed of disk material. As neutrino radiation carries away most of the thermal energy, the system eventually succumbs to its intense self-gravity and collapses inward. There are two possibilities regarding the remnant fate of the TZlO: one is that the TZlO does not collapse directly to a black hole but remains as a massive magnetar, producing a GRB akin to GRB 200826A. An SN component associated with this GRB, called an IPSN, is the result of a collision between ejecta and disk wind shell. The other is the TZlO collapses directly into a BH, and the duration of the resultant GRB is capped by the TZlO's freefall timescale, resulting in a typical short burst without subsequent IPSN components.

The event rate of WD-NS mergers is $\sim 0.5-1\times 10^{4}\rm ~Gpc^{-3} \rm yr^{-1}$\citep{Tho2009,Pas2011}, or as low as $\sim 0.7-7\times 10^{3}\rm~Gpc^{-3}yr^{-1}$ in the near cosmos \citep{Liu2018}. A non-negligible fraction of those mergers can undergo the UMT process \citep{Bob2017}, while the formation of TZlOs is contingent upon the presence of WDs with significantly high masses. WDs exceeding a mass of 1.3 $M_{\odot}$ are estimated to constitute a mere $0.5\%$ of the total WD population \citep{Kep2007}. This rarity substantially limits the formation rate of TZlOs to $\sim 50 \rm~Gpc^{-3} yr^{-1}$. Furthermore, considering the beaming effect of GRBs and applying a beaming factor of 0.01, the resultant event rate of GRBs originating from TZlO collapses is estimated to be $\sim0.035-0.5 \rm~Gpc^{-3} yr^{-1}$.

The GW emission from WD-NS mergers presents distinct characteristics compared to those from binary NS mergers and the collapse of massive stars. In the scenario of a merger involving a massive WD and an NS, the onset of mass transfer occurs as the WD fills its Roche lobe when the orbital period is approximately 3 s, corresponding to the frequency of GW emissions at around 0.1-1 Hz \citep{Kang2024}. This falls within the detection range of future space-based GW detectors, Tianqin, and LISA \citep{Luo2016}. On the other hand, we can strictly constrain the duration of stable TZlOs based on the time delay between possible GWs from WD-NS mergers and their associated GRBs.

Due to the beamed nature of GRBs, observations of coincident GWs and GRBs are not always expected in WD-NS mergers. However, optical emissions are present, particularly in scenarios involving delayed collapse. In such cases, the stellar wind shell may contain radioactive materials like $^{56}Ni$, potentially leading to low-luminosity SNe such as SNe Iax, faint rapid red transients, or Ca-rich transients \citep{Zen2020,Bob2022,Kal2023}. If the TZlO does not immediately collapse into a black hole but instead forms a massive magnetar, the interaction between the magnetar-driven ejecta and the disk wind shell can produce an IPSN. The production of an IPSN and a GRB also involves a temporal delay, marking another distinction from typical SNe associated with GRBs.

Nevertheless, discoveries of similar events in the future or the existing GRB samples can shed some light on this new class of GRB events and the fate of the WD-NS mergers.

\begin{acknowledgements}
We sincerely thank the referee for their patience and detailed suggestions, which have significantly improved this manuscript. We thank R.-F. Sheng and Y. Shao for helpful discussions about this paper. This work is supported by the National Natural Science Foundation of China (projects 12373040,12021003) and the Fundamental Research Funds for the Central Universities. We acknowledge the support by the National Key Research and Development Programs of China (2018YFA0404204, 2022YFF0711404, 2022SKA0130102), the National SKA Program of China (2022SKA0130100), the National Natural Science Foundation of China (grant Nos. 11833003, U2038105, U1831135, 12121003), the science research grants from the China Manned Space Project under NO.CMS-CSST-2021-B11, and the CNSA program D050102. The Fundamental Research Funds for the Central Universities and the Program for Innovative Talents and Entrepreneurs in Jiangsu. 

\end{acknowledgements}



\begin{thebibliography}{Fialkov \& Loeb(2017)}

\bibitem[Ahumada et al.(2020a)]{Ahu2020a} Ahumada, T., Kumar, H., Fremling, C., et al.\ 2020, GRB Coordinates Network, Circular Service, No. 29029, 29029

\bibitem[Ahumada et al.(2020b)]{Ahu2020b} Ahumada, T., Anand, S., Stein, R., et al.\ 2020, GRB Coordinates Network, Circular Service, No. 28295, 28295

\bibitem[Ahumada et al.(2021)]{Ahu2021} Ahumada, T., Singer, L.~P., Anand, S., et al.\ 2021, Nature Astronomy. doi:10.1038/s41550-021-01428-7

\bibitem[Ai et al.(2022)]{Ai2022} Ai, S., Zhang, B., \& Zhu, Z.\ 2022, \mnras, 516, 2614. doi:10.1093/mnras/stac2380

\bibitem[Alexander et al.(2020)]{Ale2020} Alexander, K.~D., Fong, W., Paterson, K., et al.\ 2020, GRB Coordinates Network, Circular Service, No. 28302, 28302

\bibitem[Althaus et al.(2005)]{Alt2005} Althaus, L.~G., Garc{\'\i}a-Berro, E., Isern, J., et al.\ 2005, \aap, 441, 689. doi:10.1051/0004-6361:20052996

\bibitem[Beaudet et al.(1967)]{Bea1967} Beaudet, G., Petrosian, V., \& Salpeter, E.~E.\ 1967, \apj, 150, 979. doi:10.1086/149398

\bibitem[Belczynski et al.(2002)]{Bel2002} Belczynski, K., Bulik, T., \& Rudak, B.\ 2002, \apj, 571, 394. doi:10.1086/339860

\bibitem[Belkin et al.(2020)]{Bel2020} Belkin, S., Zhornichenko, A., Pozanenko, A., et al.\ 2020, GRB Coordinates Network, Circular Service, No. 28306, 28306

\bibitem[Bobrick et al.(2017)]{Bob2017} Bobrick, A., Davies, M.~B., \& Church, R.~P.\ 2017, \mnras, 467, 3556. doi:10.1093/mnras/stx312

\bibitem[Bobrick et al.(2022)]{Bob2022} Bobrick, A., Zenati, Y., Perets, H.~B., et al.\ 2022, \mnras, 510, 3758. doi:10.1093/mnras/stab3574

\bibitem[Chatzopoulos et al.(2012)]{Cha2012} Chatzopoulos, E., Wheeler, J.~C., \& Vinko, J.\ 2012, \apj, 746, 121. doi:10.1088/0004-637X/746/2/121

\bibitem[D'Ai et al.(2020)]{Dai2020} D'Ai, A., Sbarufatti, B., Oates, S.~R., et al.\ 2020, GRB Coordinates Network, Circular Service, No. 28300, 28300

\bibitem[Dichiara et al.(2020)]{Dic2020} Dichiara, S., Cenko, S.~B., Troja, E., et al.\ 2020, GRB Coordinates Network, Circular Service, No. 28312, 28312

\bibitem[Eggleton(1983)]{Egg1983} Eggleton, P.~P.\ 1983, \apj, 268, 368. doi:10.1086/160960

\bibitem[Evans et al.(2007)]{Eva2007} Evans, P.~A., Beardmore, A.~P., Page, K.~L., et al.\ 2007, \aap, 469, 379. doi:10.1051/0004-6361:20077530

\bibitem[Evans et al.(2009)]{Eva2009} Evans, P.~A., Beardmore, A.~P., Page, K.~L., et al.\ 2009, \mnras, 397, 1177. doi:10.1111/j.1365-2966.2009.14913.x

\bibitem[Fern{\'a}ndez et al.(2019)]{Fer2019} Fern{\'a}ndez, R., Margalit, B., \& Metzger, B.~D.\ 2019, \mnras, 488, 259. doi:10.1093/mnras/stz1701

\bibitem[Gao et al.(2022)]{Gao2022} Gao, H., Lei, W.-H., \& Zhu, Z.-P.\ 2022, \apjl, 934, L12. doi:10.3847/2041-8213/ac80c7

\bibitem[Gottlieb et al.(2023)]{Got2023} Gottlieb, O., Metzger, B.~D., Quataert, E., et al.\ 2023, \apjl, 958, L33. doi:10.3847/2041-8213/ad096e

\bibitem[Hotokezaka et al.(2013)]{Hot2013} Hotokezaka, K., Kiuchi, K., Kyutoku, K., et al.\ 2013, \prd, 87, 024001. doi:10.1103/PhysRevD.87.024001

\bibitem[Huang et al.(2000)]{Hua2000} Huang, Y.~F., Dai, Z.~G., \& Lu, T.\ 2000, \mnras, 316, 943. doi:10.1046/j.1365-8711.2000.03683.x


\bibitem[Kasen \& Bildsten(2010)]{Kas2010} Kasen, D. \& Bildsten, L.\ 2010, \apj, 717, 245. doi:10.1088/0004-637X/717/1/245

\bibitem[Kaltenborn et al.(2023)]{Kal2023} Kaltenborn, M.~A.~R., Fryer, C.~L., Wollaeger, R.~T., et al.\ 2023, \apj, 956, 71. doi:10.3847/1538-4357/acf860

\bibitem[Kang et al.(2024)]{Kang2024} Kang, Y., Liu, C., Zhu, J.-P., et al.\ 2024, \mnras, 528, 5309. doi:10.1093/mnras/stae340

\bibitem[Kepler et al.(2007)]{Kep2007} Kepler, S.~O., Kleinman, S.~J., Nitta, A., et al.\ 2007, \mnras, 375, 1315. doi:10.1111/j.1365-2966.2006.11388.x

\bibitem[Klu{\'z}niak \& Ruderman(1998)]{Klu1998} Klu{\'z}niak, W. \& Ruderman, M.\ 1998, \apjl, 505, L113. doi:10.1086/311622

\bibitem[Kotera et al.(2013)]{Kot2013} Kotera, K., Phinney, E.~S., \& Olinto, A.~V.\ 2013, \mnras, 432, 3228. doi:10.1093/mnras/stt680


\bibitem[Lasky et al.(2017)]{Las2017} Lasky, P.~D., Leris, C., Rowlinson, A., et al.\ 2017, \apjl, 843, L1. doi:10.3847/2041-8213/aa79a7


\bibitem[Liu(2018)]{Liu2018} Liu, X.\ 2018, \apss, 363, 242. doi:10.1007/s10509-018-3462-3

\bibitem[Luo et al.(2016)]{Luo2016} Luo, J., Chen, L.-S., Duan, H.-Z., et al.\ 2016, Classical and Quantum Gravity, 33, 035010. doi:10.1088/0264-9381/33/3/035010

\bibitem[L{\"u} et al.(2014)]{Lv2014} L{\"u}, H.-J., Zhang, B., Liang, E.-W., et al.\ 2014, \mnras, 442, 1922. doi:10.1093/mnras/stu982

\bibitem[Margalit \& Metzger(2016)]{Mar2016} Margalit, B. \& Metzger, B.~D.\ 2016, \mnras, 461, 1154. doi:10.1093/mnras/stw1410


\bibitem[Metzger(2012)]{Met2012} Metzger, B.~D.\ 2012, \mnras, 419, 827. doi:10.1111/j.1365-2966.2011.19747.x

\bibitem[Middleditch(2004)]{Mid2004} Middleditch, J.\ 2004, \apjl, 601, L167. doi:10.1086/382074

\bibitem[Nauenberg(1972)]{Nau1972} Nauenberg, M.\ 1972, \apj, 175, 417. doi:10.1086/151568

\bibitem[Paschalidis et al.(2009)]{Pas2009} Paschalidis, V., MacLeod, M., Baumgarte, T.~W., et al.\ 2009, \prd, 80, 024006. doi:10.1103/PhysRevD.80.024006

\bibitem[Paschalidis et al.(2011)]{Pas2011} Paschalidis, V., Liu, Y.~T., Etienne, Z., et al.\ 2011, \prd, 84, 104032. doi:10.1103/PhysRevD.84.104032

\bibitem[Rastinejad et al.(2022)]{Ras2022} Rastinejad, J.~C., Gompertz, B.~P., Levan, A.~J., et al.\ 2022, \nat, 612, 223. doi:10.1038/s41586-022-05390-w

\bibitem[Rhodes et al.(2021)]{Rho2021} Rhodes, L., Fender, R., Williams, D.~R.~A., et al.\ 2021, \mnras, 503, 2966. doi:10.1093/mnras/stab640

\bibitem[Rossi et al.(2020)]{Ros2020} Rossi, A., D'Avanzo, P., D'Elia, V., et al.\ 2020, GRB Coordinates Network, Circular Service, No. 28949, 28949

\bibitem[Rossi et al.(2022)]{Ros2022} Rossi, A., Rothberg, B., Palazzi, E., et al.\ 2022, \apj, 932, 1. doi:10.3847/1538-4357/ac60a2

\bibitem[Rothberg et al.(2020)]{Rot2020} Rothberg, B., Kuhn, O., Veillet, C., et al.\ 2020, GRB Coordinates Network, Circular Service, No. 28319, 28319

\bibitem[Ryan et al.(2020)]{Rya2020} Ryan, G., van Eerten, H., Piro, L., et al.\ 2020, \apj, 896, 166. doi:10.3847/1538-4357/ab93cf

\bibitem[Sun et al.(2023)]{Sun2023} Sun, H., Wang, C.-W., Yang, J., et al.\ 2023, arXiv:2307.05689. doi:10.48550/arXiv.2307.05689

\bibitem[Tanaka et al.(2020)]{Tan2020} Tanaka, M., Kato, D., Gaigalas, G., et al.\ 2020, \mnras, 496, 1369. doi:10.1093/mnras/staa1576

\bibitem[Thompson et al.(2009)]{Tho2009} Thompson, T.~A., Kistler, M.~D., \& Stanek, K.~Z.\ 2009, arXiv:0912.0009

\bibitem[Th{\"o}ne et al.(2011)]{Tho2011} Th{\"o}ne, C.~C., de Ugarte Postigo, A., Fryer, C.~L., et al.\ 2011, \nat, 480, 72. doi:10.1038/nature10611

\bibitem[Thorne \& Zytkow(1977)]{Tho1977} Thorne, K.~S. \& Zytkow, A.~N.\ 1977, \apj, 212, 832. doi:10.1086/155109

\bibitem[Toonen et al.(2018)]{Too2018} Toonen, S., Perets, H.~B., Igoshev, A.~P., et al.\ 2018, \aap, 619, A53. doi:10.1051/0004-6361/201833164

\bibitem[Troja et al.(2022)]{Tro2022} Troja, E., Fryer, C.~L., O'Connor, B., et al.\ 2022, \nat, 612, 228. doi:10.1038/s41586-022-05327-3

\bibitem[Verbunt \& Rappaport(1988)]{Ver1988} Verbunt, F. \& Rappaport, S.\ 1988, \apj, 332, 193. doi:10.1086/166645

\bibitem[Wang et al.(2022)]{Wang2022} Wang, X.~I., Zhang, B.-B., \& Lei, W.-H.\ 2022, \apjl, 931, L2. doi:10.3847/2041-8213/ac6c7e

\bibitem[Wang et al.(2024)]{Wang2024} Wang, X.~I., Yu, Y.-W., Ren, J., et al.\ 2024, \apjl, 964, L9. doi:10.3847/2041-8213/ad2df6

\bibitem[Wheeler et al.(1998)]{Whe1998} Wheeler, J.~C., Cowan, J.~J., \& Hillebrandt, W.\ 1998, \apjl, 493, L101. doi:10.1086/311133

\bibitem[Yang et al.(2020)]{Yan2020} Yang, J., Chand, V., Zhang, B.-B., et al.\ 2020, \apj, 899, 106. doi:10.3847/1538-4357/aba745

\bibitem[Yang et al.(2022)]{Yang2022} Yang, J., Ai, S., Zhang, B.-B., et al.\ 2022, \nat, 612, 232. doi:10.1038/s41586-022-05403-8

\bibitem[Yang et al.(2024)]{Yang2024} Yang, Y.-H., Troja, E., O'Connor, B., et al.\ 2024, \nat, 626, 742. doi:10.1038/s41586-023-06979-5


\bibitem[Yi \& Blackman(1998)]{Yi1998} Yi, I. \& Blackman, E.~G.\ 1998, \apjl, 494, L163. doi:10.1086/311192

\bibitem[Yu et al.(2013)]{Yu2013} Yu, Y.-W., Zhang, B., \& Gao, H.\ 2013, \apjl, 776, L40. doi:10.1088/2041-8205/776/2/L40

\bibitem[Yu et al.(2018)]{Yu2018} Yu, Y.-W., Liu, L.-D., \& Dai, Z.-G.\ 2018, \apj, 861, 114. doi:10.3847/1538-4357/aac6e5

\bibitem[Zhang(2011)]{Zha2011} Zhang, B.\ 2011, Comptes Rendus Physique, 12, 206. doi:10.1016/j.crhy.2011.03.004


\bibitem[Zhang et al.(2015)]{Zha2015} Zhang, B.-B., van Eerten, H., Burrows, D.~N., et al.\ 2015, \apj, 806, 15. doi:10.1088/0004-637X/806/1/15

\bibitem[Zhang et al.(2021)]{Zha2021} Zhang, B.-B, Liu, Z.-K, Peng, Z.-K., et al. Nat Astron (2021). doi:10.1038/s41550-021-01395-z

\bibitem[Zhang \& Dai(2009)]{Zhang2009} Zhang, D. \& Dai, Z.~G.\ 2009, \apj, 703, 461. doi:10.1088/0004-637X/703/1/461

\bibitem[Zhang \& Dai(2010)]{Zhang2010} Zhang, D. \& Dai, Z.~G.\ 2010, \apj, 718, 841. doi:10.1088/0004-637X/718/2/841

\bibitem[Zenati et al.(2019)]{Zen2019} Zenati, Y., Perets, H.~B., \& Toonen, S.\ 2019, \mnras, 486, 1805. doi:10.1093/mnras/stz316

\bibitem[Zenati et al.(2020)]{Zen2020} Zenati, Y., Bobrick, A., \& Perets, H.~B.\ 2020, \mnras, 493, 3956. doi:10.1093/mnras/staa507

\bibitem[Zhong \& Dai(2020)]{Zho2020} Zhong, S.-Q. \& Dai, Z.-G.\ 2020, \apj, 893, 9. doi:10.3847/1538-4357/ab7bdf

\bibitem[{{Zhong} {et~al.}(2023){Zhong}, {Li}, \& {Dai}}]{Zhong2023}
{Zhong}, S.-Q., {Li}, L., \& {Dai}, Z.-G. 2023, \apjl, 947, L21, doi:10.3847/2041-8213/acca83

\end{thebibliography}
\end{document}